# Nanolamellar Hybrid High-Entropy Alloys with Superior Micromechanical Properties

Shivam Dangwal[1,2], Yoji Mine[3], Shohei Ueki[4], Xavier Sauvage[5], Fabien Cuvilly[5], Liliana Romero Resendiz[6], Muhammad Naeem[6], Kaveh Edalati[1,2,*]

[1] WPI, International Institute for Carbon-Neutral Energy Research (WPI-I2CNER), Kyushu University, Fukuoka 819-0395, Japan
[2] Department of Automotive Science, Graduate School of Integrated Frontier Sciences, Kyushu University, Fukuoka 819-0395, Japan
[3] Department of Materials Science and Engineering, Kumamoto University, Kumamoto 860-8555, Japan
[4] Department of Mechanical Engineering, Kyushu University, 744 Motooka, Nishi-ku, Fukuoka 819-0395, Japan
[5] Univ Rouen Normandie, INSA Rouen Normandie, CNRS, Groupe de Physique des Matériaux, UMR6634, 76000 Rouen, France
[6] Interdisciplinary Research Center, Liaoning Academy of Materials, Shenyang 110167, China

Metallic materials with nanolamellar structures, such as pearlitic steels, exhibit high strength with appropriate ductility. Considering the potential ability of such nanolamellar structures to break the traditional strength-ductility trade-off in metallic alloys, this study aims at developing a unique nanolamellar structure with superior micromechanical properties by combining two different high-entropy alloys (HEAs). $Al_{0.1}CoCrFeNi$ with the face-centered cubic (FCC) structure is combined with TiZrHfNbTa with the body-centered cubic (BCC) structure using high-pressure torsion (HPT) of half discs of each alloy. That way, a layered hybrid structure was formed, with layer thickness down to about 61 nm. The BCC/FCC nanolamellar hybrid structure exhibits an exceptional combination of properties with an ultimate tensile strength of 2.4 GPa, a maximum bending strength of 4.0 GPa, and a hardness of 740 Hv, while retaining some ductility/plasticity under both tensile and bending loads. Detailed analyses by synchrotron diffraction, electron microscopy and atom probe tomography suggests that these high strength and hardness, which are superior to those of nanostructured HEAs, result from: (i) extreme grain boundary strengthening from nanograins with a mean size of 22 nm, (ii) presence of defects such as dislocations in FCC and BCC, stacking faults in FCC and twins in FCC, and (iii) interphase hardening from BCC/FCC nanolamellar boundaries with about 30% contribution to the total hardness. This work demonstrates that combining two HEAs using HPT into a defect-rich hybrid nanolamellar composite forms a promising synergy of ultrahigh strength and reasonable ductility/plasticity.

*Corresponding author (E-mail: kaveh.edalati@kyudai.jp; Tel/Fax: +81 92 802 6744)

## Introduction

The design of structural materials having high strength and reasonable plasticity before failure has been a major task for materials scientists. Strength is the ability of materials to resist deformation, whereas ductility is the capability of materials to bear plastic deformation [1]. As these definitions suggest, one resists the change, and the other allows it, leading to a fundamental trade-off between them. Overcoming this inherent trade-off has been a key scientific and engineering issue in developing high-strength materials. The strength of materials can be enhanced by mechanisms developed in the last century, such as work hardening via dislocations [2], grain size reduction [3] and precipitation introduction [4]; however, these hardening processes can lead to poor ductility. Some studies suggested that combining or modifying different mechanisms, such as ultrafine-grain refinement along with nano twinning [5,6], introducing lamellar microstructure [7,8], controlled precipitation hardening [9,10], etc., can improve strain hardening and moderate the inverse strength-ductility relation.

The development of novel alloys capable of the simultaneous activation of multiple hardening/deformation mechanisms is a key point that can be used to acquire high-strength materials while retaining some ductility. A class of materials named high-entropy alloys (HEAs) has arisen as a potential material class to implement different hardening/deformation mechanisms simultaneously. HEAs are alloys that are made by mixing five or more elements such that their mixing entropy is over 1.5$R$ ($R$: the gas constant) [11,12]. The coexistence of several elements in HEAs provides them with unique compositional complexity and structural tunability. Although HEAs have been explored for their application in various fields such as hydrogen storage [13-15], photocatalysis [16,17], superconductors [18,19], biomaterials [20,21], refractory materials [22-24], etc., the highest interest in these alloys is still due to their promising mechanical properties [25]. While the unique chemical complexity of HEAs provides a significant advantage to beat the inverse strength-ductility relation, further microstructural engineering is required to attain high strength in these alloys.

A promising processing route for microstructural engineering of metallic alloys is via severe plastic deformation (SPD) [26,27]. SPD techniques like multi-directional forging (MDF) [28], accumulative roll-bonding (ARB) [29], equal-channel angular pressing (ECAP) [30], twist extrusion (TE) [31] and high-pressure torsion (HPT) [32], can successfully form nanostructured materials, while some continuous SPD methods have been commercialized in recent years [33]. Among SPD processes, HPT is a powerful method that simultaneously imposes extremely high pressure and torsional shear strain in materials [28], as depicted in Fig. 1a. The HPT technology is not only the most effective in grain refinement, but also promotes atomic diffusion [34] and solid-state reactions [24,35], increases dislocation density even within nanograin [36], forms twin boundaries in alloys with low stacking faults [37], and accordingly generates high strength and hardness [38].

The application of HPT has been explored beyond microstructural tailoring to the synthesis of various new materials and to design hybrid materials containing dissimilar metals [39,40]. While the main objective of using HPT to hybrid compounds is mixing dissimilar materials, this processing can occasionally form intermetallics and/or non-equilibrium phases [41-43]. One potential function of HPT, which has largely been overlooked in the literature, is the generation of layered composites by processing half discs of two different materials [41]. While ARB can fabricate layered composites, such as Cu/Ni/Al [44], Al/Cu/Sn/Ni [45], Al-1060/Al-7N01 [46], Al/Cu/Zn/Ni [47] and Cu/CuAg/NbMoTaW [48], but multiple rolling and stacking are necessary in ARB, while HPT can produce such layered structures with one continuous process. Such layered architectures occasionally form in nature, such as in bone and bamboo, and are present in some alloys like pearlitic steels, which can show a promising mixture of strength and toughness [49-51].

Within the HEA family, some studies have introduced lamellar structures in HEAs through casting [52,53], directional solidification [54], additive manufacturing [53], or thermo-mechanical processing of single- or dual-phase alloys [55,56], but these approaches typically rely on *in situ* lamellae formed from a single parent alloy composition, usually in eutectic systems. Layered architectures combining face-centered cubic (FCC) and body-centered cubic (BCC) phases have also been pursued through nanoscale multilayer thin films [57], spark plasma sintering [58] and HPT processing of full discs stacked on each other [59]. However, HPT processing of half discs of BCC and FCC HEAs, which is principally simpler than stacking many layers on each other and can geometrically produce fine layers by just increasing HPT turns, remains unexplored.

Inspired by the promising mechanical properties of HEAs as well as layered structures, this study aims to engineer a layered structure made up of two HEAs ($Al_{0.1}CoCrFeNi$ and TiZrHfNbTa) through HPT. These two HEAs are selected because $Al_{0.1}CoCrFeNi$ has an FCC phase and TiZrHfNbTa has a BCC phase. $Al_{0.1}CoCrFeNi$ exhibits good ductility even at cryogenic temperatures [60,61], while TiZrHfNbTa shows high strength [25,62,63]. To the authors' knowledge, the combination of HEAs FCC $Al_{0.1}CoCrFeNi$ and BCC TiZrHfNbTa has not been explored as a hybrid material by any fabrication route, including HPT, and the microstructure and micromechanical properties of such a hybrid material remain unexplored. The present work introduces HPT of half discs as an effective fabrication route for hybrid nanolamellar structures composed of two dissimilar HEAs. The hybrid $Al_{0.1}CoCrFeNi$ + TiZrHfNbTa material processed through HPT exhibits a nanolamellar configuration with a mean grain size of 22 nm, and its strength under tension and bending reaches 2.4 and 4.0 GPa, respectively, while retaining some ductility. These results introduce nanolamellar hybrid HEAs produced by HPT as potential materials with excellent mechanical properties.

## Experimental Procedures

Ingots of HEA $Al_{0.1}CoCrFeNi$ and TiZrHfNbTa were fabricated separately via arc melting using aluminum (99.99%), cobalt (99.9%), chromium (99.99%), iron (99.9%), nickel (99.99%), titanium (99.9%), zirconium (99.2%), hafnium (99.7%), niobium (99.9%) and tantalum (99.9%). The ingots of each alloy were synthesized separately using a copper crucible with water circulation in a protective atmosphere of argon. Prior to HPT processing, the ingots of $Al_{0.1}CoCrFeNi$ and TiZrHfNbTa were cut into semicircular discs, having a 5 mm radius and 0.80 mm thickness, employing a wire-cutting electric discharge machine. The HPT process was carried out on semicircular discs of these two HEAs, which were put next to each other between two anvils (Fig. 1b). The processing rotation speed, pressure and temperature were $\omega = 1$ rpm, $P = 6$ GPa and $T = 300$ K, respectively. By applying the rotation numbers ($N$) as 1, 10 and 50, strain was manipulated during HPT. A photograph of the hybrid HEA $Al_{0.1}CoCrFeNi$ + TiZrHfNbTa, following the HPT process for one rotation, is depicted in Fig. 1c, suggesting good connectivity between these two HEAs. The good connectivity should be due to high pressure and strain and not temperature rise, because the temperature rise was as small as 20 K after 50 turns, which is consistent with previous publications on HPT processing of metals [64,65] and ultra-hard HEAs [66]. After HPT processing, various methods were used to characterize the materials and assess their mechanical properties.

To assess the crystallographic features, X-ray diffraction (XRD) of HPT-processed materials was carried out. A 45 kV voltage and a 200 mA filament current were employed to generate the Cu Kα X-ray radiation. The software PDXL was used to perform Rietveld analysis for determining crystal structure and lattice constants.

The diffraction patterns for material processed with $N$ = 50 were obtained using synchrotron high-energy X-ray diffraction (HEXRD) at Diamond Light Source using beamline I12-JEEP. The HEXRD analysis was done with a beam energy of 78.234 keV (wavelength of 0.15848 Å), and a beam size of 0.5×0.5 $mm^2$. The HEXRD data were acquired utilizing a Pilatus 2M CdTe detector located at 601.8 mm from the specimen. A transmission geometry was used for HEXRD analyses, while the exposure time was 4 s per diffraction pattern. The acquired 2D diffraction rings were azimuthally integrated using DAWN software to obtain 1D diffraction profiles. A $CeO_2$ pattern measured under identical conditions was used as the reference to perform energy and geometry calibration. The line-profile measurements were made at radial distances from $r$ = 0.5 mm to $r$ = 4.5 mm at every 0.5 mm on the disc.

To evaluate the hardness of the hybrid materials, a bar with 10 mm length and a 0.7×0.7 $mm^2$ cross-section was extracted from the central region of HPT-deformed discs by utilizing a wire-cutting electric discharge machine. The cross-sectional face was first ground with different emery papers and finally polished with alumina powder and a buff to produce a mirror-like surface. Vickers hardness testing was done on the cross-sectional face with a force of 0.3 kgf.

Microstructural and compositional analyses at the micrometer scale were conducted using field-emission scanning electron microscopy (SEM) and energy-dispersive X-ray spectroscopy (EDS) with an acceleration voltage of 15 kV. For SEM-EDS analysis, the cross-sectional face of the bar, which was prepared for the Vickers microhardness test, was first ground with emery papers 2000 grit, and then polished with two diamond suspensions (9 µm and 3 µm sizes, respectively) and subsequently with colloidal silica nanopowders (60 nm size). SEM-EDS images were obtained at 2.5 mm from the specimen center.

Nanostructural and chemical analyses at the nanometer level for the specimen deformed by HPT with $N$ = 50 were conducted by transmission and scanning-transmission electron microscopes (TEM and STEM). A wire-cutting electric discharge machine was employed for cutting a cylindrical disc of 1.5 mm radius at 1-4 mm from the center of the specimen. This disc was thinned to 100 µm using sandpapers and electropolished using a mixture of 5 vol% perchloric acid, 30 vol% butanol and 65 vol% methanol at 263 K by applying 20 V voltage. A precision ion polishing system was used for ion milling after electropolishing. Ion milling was conducted with an argon ion flow having an energy of 5 keV for 40 minutes with incident angles of ±5º. An additional TEM sample was also produced via a focused ion beam (FIB) process to have a cross-sectional view. Selected area electron diffraction (SAED), TEM bright- and dark-field images, high-resolution images, high-angle annular dark-field (HAADF) images and EDS mappings were utilized for the analyses. Fast Fourier transform (FFT) was employed to assess high-resolution images.

To analyze the atomic-scale distribution of elements, the specimen deformed by HPT for $N$ = 50 was tested by atom probe tomography (APT). APT specimens were extracted via FIB from 2.5 mm from the specimen center. APT analyses were performed utilizing a CAMECA LEAP-4000HR instrument at $10^{-9}$ Pa vacuum at 50 K temperature. A 20 pJ laser with 200 kHz pulse frequency was used for evaporating and ionizing the sample, while the pulse fraction was 20%. IVAS (Integrated Visualization & Analysis Software) was used to reconstruct the analyzed volume, and GPM3DSoft (IDDN.FR.001.430017.000.S.P.2020.000.10000) was used for analyzing the APT data.

To assess mechanical properties, micro-tensile and micro-bending tests were employed. For these micromechanical tests, FIB was used to prepare miniature samples, with dimensions given in Fig. 1d and 1e, from the specimen treated by HPT with $N = 50$. For the micro-tensile test, a sample with dimensions of 20×20×50 $\mu m^3$ was fabricated using FIB. The loading axis for the tensile specimen was perpendicular to the surface of the disc, and its position from the disc center was 2.5 mm. The tensile test was done at ambient temperature at a speed of 0.1 $\mu m\ s^{-1}$, which is equivalent to a strain rate of $2\times10^{-3}\ s^{-1}$. For the bending test, a micro-cuboid with dimensions of 15×15×40 $\mu m^3$ was prepared from 2.5 mm from the specimen center. The beams were arranged in a cantilever configuration, with one end fixed to the bulk sample. The other end was subjected to loading via a nanoindentation device. The micro-bending test was performed at ambient temperature perpendicular to the disc surface with a load cell having a maximum capacity of 200 mN. The stress was estimated according to Euler-Bernoulli beam theory for bending, while the strain was reported as bending displacement [1,67].

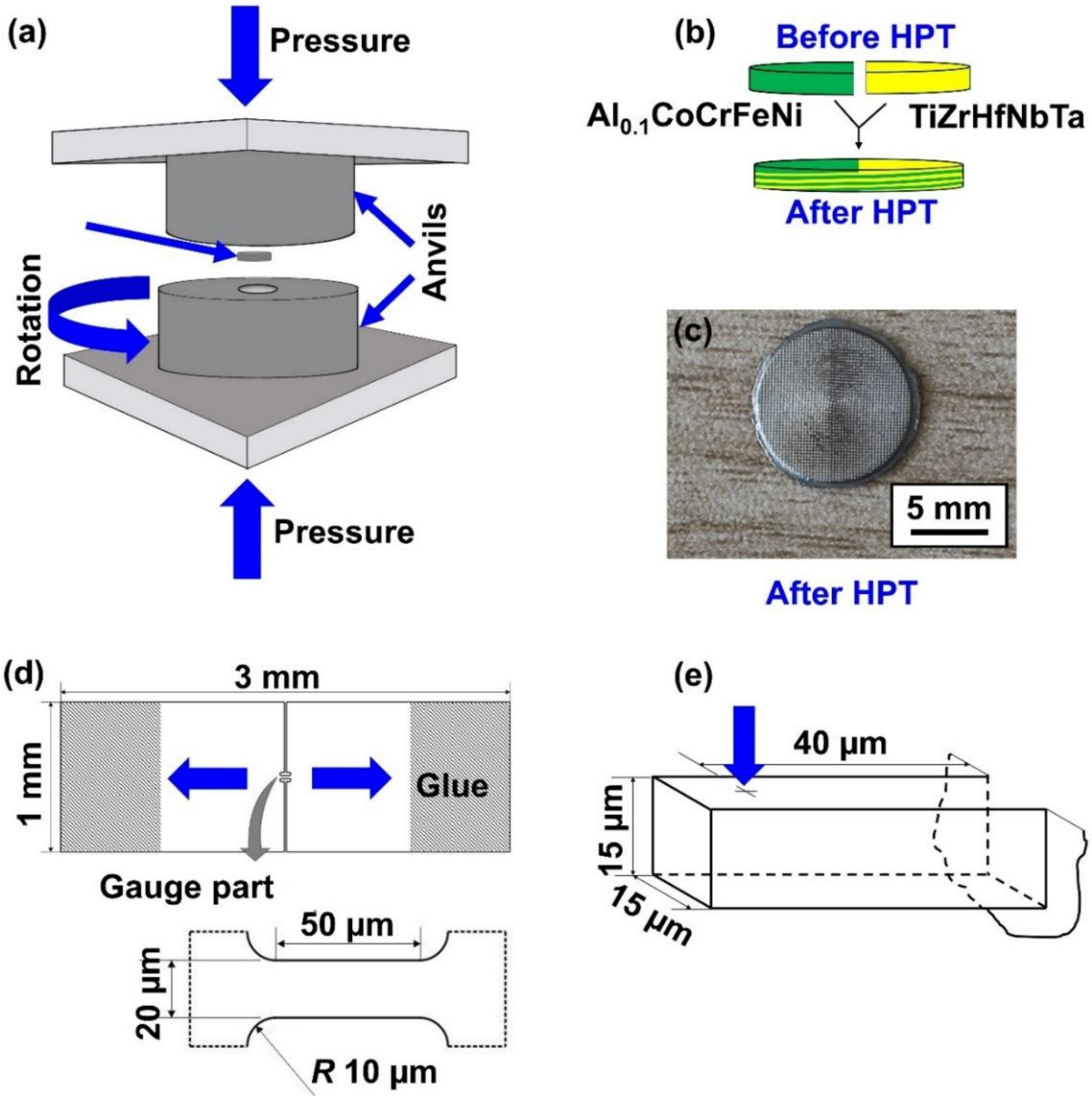

Figure 1. (a) Schematic depiction of high-pressure torsion (HPT), in which high hydrostatic pressure is applied through two anvils to a disc sample with applying shear strain by simultaneous rotation of anvils. (b) Schematic of HEA $Al_{0.1}CoCrFeNi$ + TiZrHfNbTa semicircular discs before and after HPT processing. (c) Photograph of two semicircular discs after HPT processing for $N = 1$ turn. Dimensions of samples for (d) micro-tensile test and (e) micro-bending test.

## Results

XRD profiles of the hybrid material deformed via HPT with $N = 1$, 10 and 50 are depicted in Fig. 2a. All three processed hybrid HEAs show the presence of FCC and BCC structures. This agrees with the initial FCC phase of HEA $Al_{0.1}CoCrFeNi$ [60,61] and the BCC phase of HEA TiZrHfNbTa [25,62,63], suggesting that no phase transition takes place during the HPT treatment. The results of Rietveld refinement for hybrid HEA $Al_{0.1}CoCrFeNi$ + TiZrHfNbTa deformed with $N = 50$ are depicted in Fig. 2b. An excellent agreement between measured and calculated patterns is confirmed by a low weighted profile residual ($R_{wp} = 4.21$ %). The lattice parameters for FCC are $a = b = c = 0.358$ nm, and for BCC are $a = b = c = 0.342$ nm, which do not change with HPT turns, as presented in Table 1. The insignificant change in the lattice parameters suggests a minor mechanical alloying between the two HEAs during HPT processing.

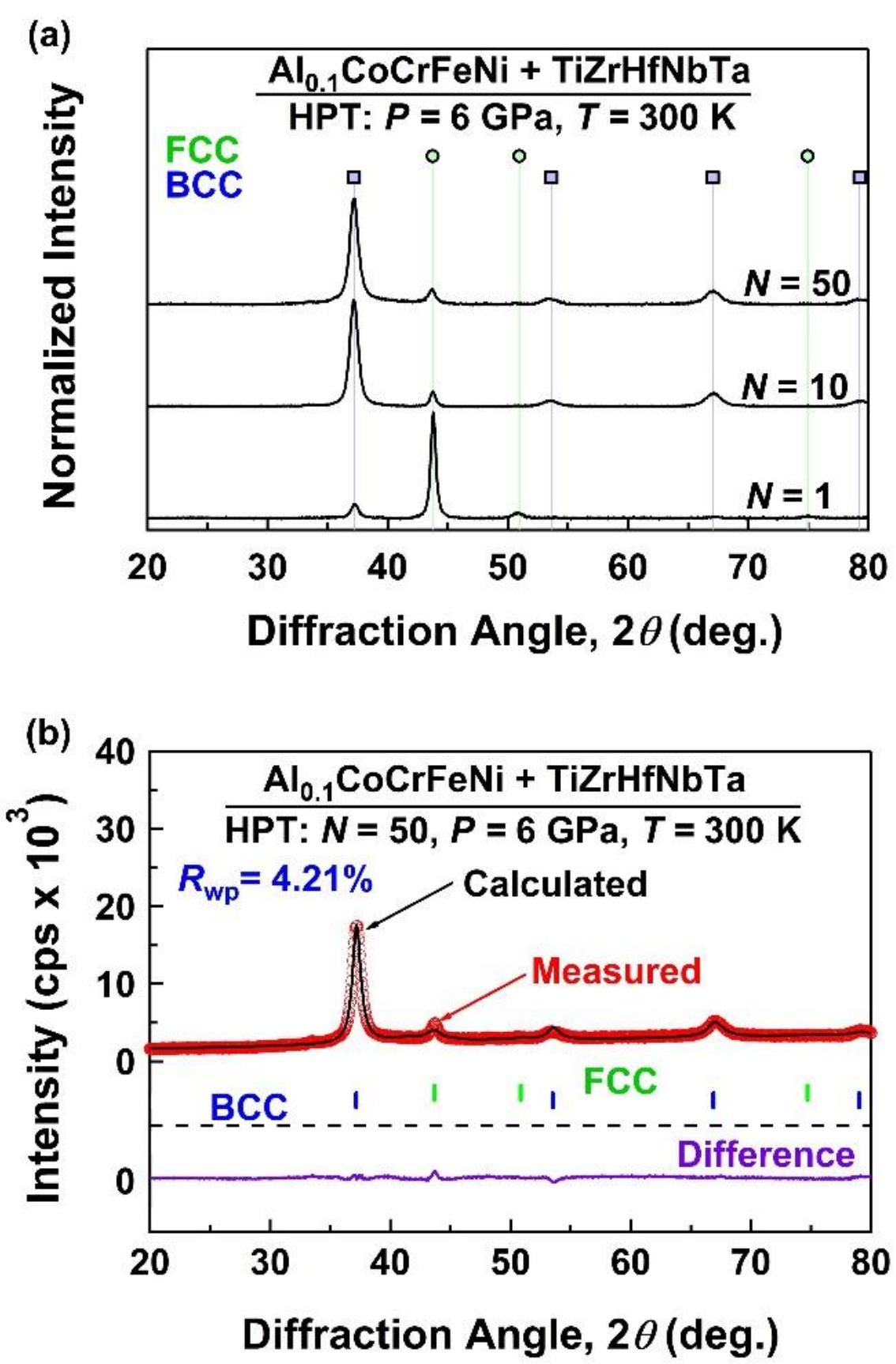

Figure 2. Hybrid high-entropy alloy $Al_{0.1}CoCrFeNi$ + TiZrHfNbTa deformed via high-pressure torsion (HPT), showing mixed face-centered cubic (FCC) and body-centered cubic (BCC) structures. (a) X-ray diffraction of specimens deformed with $N = 1$, 10 and 50 HPT rotations, and (b) XRD profile and related Rietveld refinement for specimen deformed with $N = 50$ HPT rotations.

Table 1. Lattice constants and compositional quantification of phases in hybrid high-entropy alloy $Al_{0.1}$CoCrFeNi + TiZrHfNbTa deformed via high-pressure torsion (HPT) with $N$ = 1, 10 and 50 rotations.

| | Nominal | | $N$ = 1 | | $N$ = 10 | | $N$ = 50 | |
|---|---|---|---|---|---|---|---|---|
| | **FCC** | **BCC** | **FCC** | **BCC** | **FCC** | **BCC** | **FCC** | **BCC** |
| **Lattice Parameters, *a* (nm)** | | | 0.358 | 0.342 | 0.358 | 0.342 | 0.358 | 0.342 |
| **Al (at%)** | 2.4 | 0.0 | 2.3 | 0.3 | 2.7 | 0.0 | 2.6 | 0.1 |
| **Co (at%)** | 24.4 | 0.0 | 21.4 | 0.0 | 22.1 | 0.8 | 21.3 | 0.1 |
| **Cr (at%)** | 24.4 | 0.0 | 25.5 | 0.2 | 25.7 | 0.7 | 25.0 | 0.4 |
| **Fe (at%)** | 24.4 | 0.0 | 25.6 | 0.7 | 25.3 | 0.9 | 25.9 | 0.3 |
| **Ni (at%)** | 24.4 | 0.0 | 25.1 | 0.0 | 23.8 | 0.9 | 24.3 | 0.2 |
| **Ti (at%)** | 0.0 | 20.0 | 0.0 | 19.6 | 0.0 | 18.1 | 0.2 | 19.4 |
| **Zr (at%)** | 0.0 | 20.0 | 0.1 | 19.6 | 0.1 | 19.0 | 0.2 | 19.1 |
| **Hf (at%)** | 0.0 | 20.0 | 0.0 | 20.4 | 0.0 | 22.5 | 0.1 | 20.1 |
| **Nb (at%)** | 0.0 | 20.0 | 0.0 | 20.4 | 0.1 | 20.0 | 0.2 | 20.7 |
| **Ta (at%)** | 0.0 | 20.0 | 0.0 | 18.8 | 0.2 | 17.1 | 0.2 | 19.6 |

Synchrotron diffraction patterns of the hybrid specimen deformed with $N$ = 50 are shown in Fig. 3 at different radial locations from the specimen center. The coexistence of only FCC and BCC crystals is observed in the patterns, indicating the absence of phase transformation throughout the disc. Moreover, no peak shifts are observed by moving away from the specimen center (i.e., rising shear strain), suggesting the absence of significant mechanical alloying by HPT processing. A slight peak broadening toward the disc edge suggests more significant microstructural refinement with increasing shear strain [27].

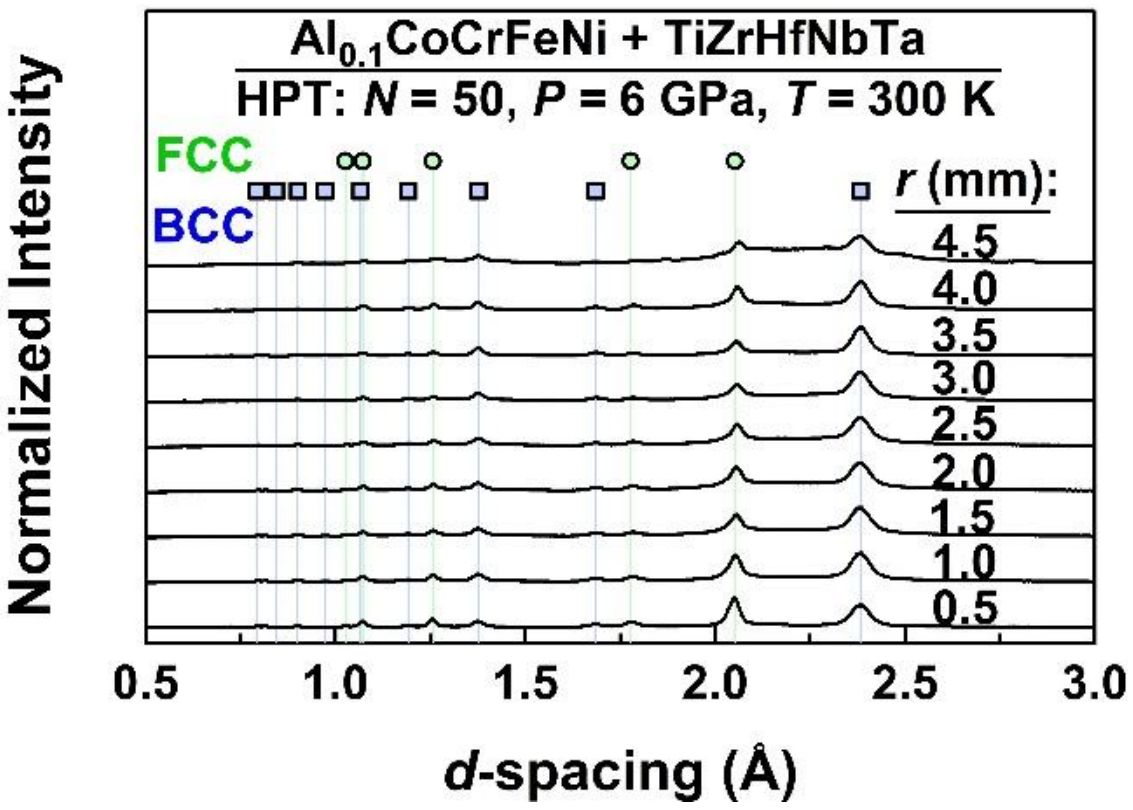

Figure 3. Mixed FCC and BCC structures across the radial direction in hybrid high-entropy alloy $Al_{0.1}$CoCrFeNi + TiZrHfNbTa deformed via high-pressure torsion (HPT). Synchrotron high-energy X-ray diffraction patterns in sequential radial positions ($r$) of specimen deformed with $N$ = 50 HPT rotations.

SEM backscatter-electron images and relevant EDS elemental maps of the hybrid materials deformed for $N$ = 1, 10 and 50 are depicted in Fig. 4, where the shear direction is perpendicular to the images. A layered structure is observed in the SEM images, containing bright and dark phases. The dark phase corresponds to lighter elements, which are in the FCC phase of $Al_{0.1}$CoCrFeNi, while the bright phase corresponds to heavier elements, which are in the BCC phase of TiZrHfNbTa. SEM-EDS images further confirm the elemental distribution in each sample. Quantified elemental composition of hybrid HEA $Al_{0.1}$CoCrFeNi + TiZrHfNbTa is shown in Table 1, indicating the composition of phases is consistent with the nominal compositions. Table 1 also suggests that overall mechanical alloying between the two phases is negligible even after $N = 50$ turns. It should be noted that the data in Table 1 were obtained from layers with larger sizes than the spatial resolution of SEM-EDS. With $N = 1$, the formation of three layers can be detected in the SEM images. With a higher number of turns, this mixing is enhanced, thereby forming more layers with smaller heights down to the nanometer level after $N = 50$.

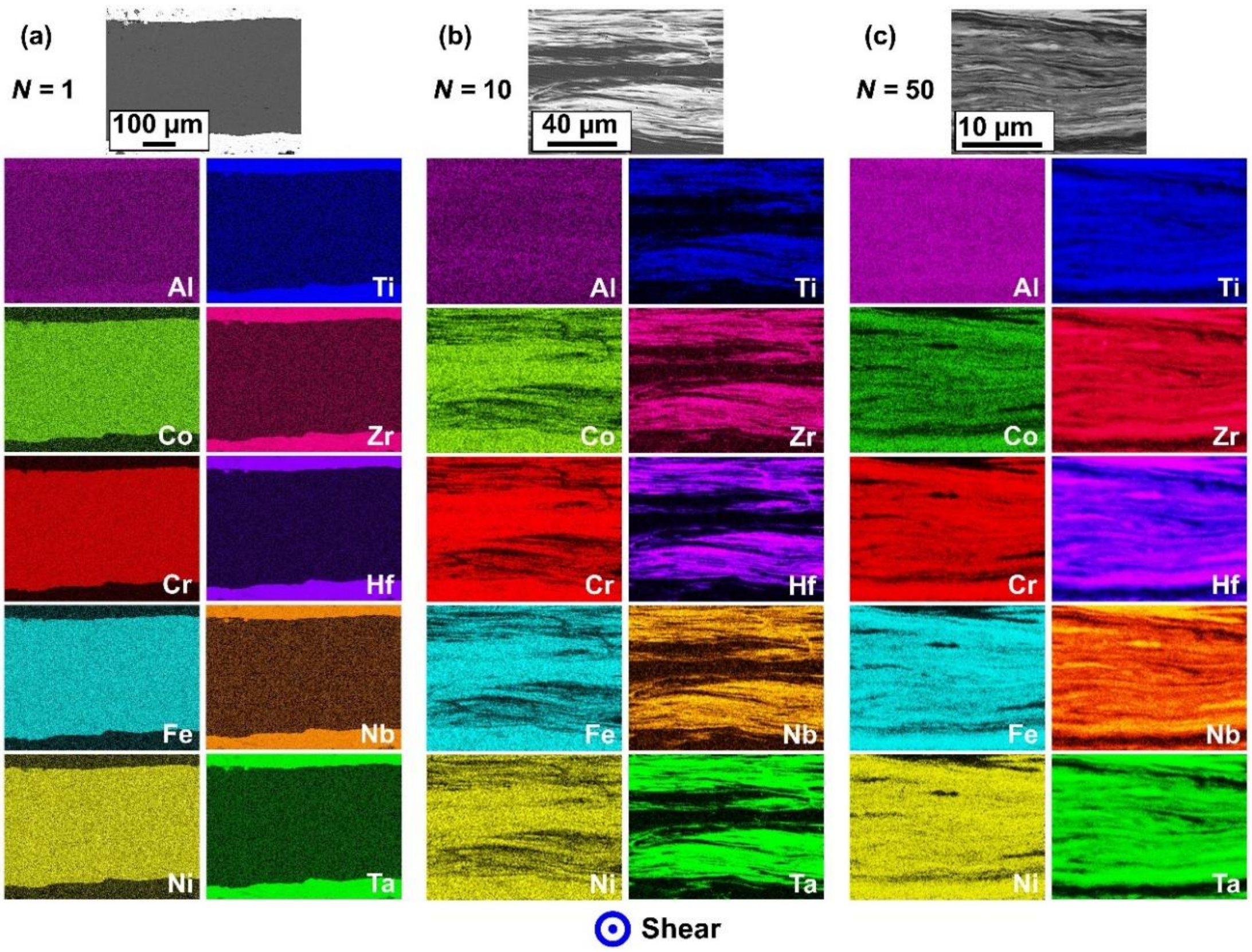

Figure 4. Formation of layered configuration in hybrid high-entropy alloy $Al_{0.1}CoCrFeNi$ + TiZrHfNbTa deformed via high-pressure torsion (HPT). Scanning electron microscopy images from cross section in radial direction (top) and relevant elemental maps by energy-dispersive X-ray spectroscopy (bottom) for specimens deformed with (a) $N = 1$, (b) $N = 10$ and (c) $N = 50$ HPT rotations.

Since the thickness of layers after $N = 50$ was too small in many regions for SEM-EDS, STEM-EDS was conducted to confirm the formation of a layered structure. HAADF and corresponding STEM-EDS for the specimen deformed with $N = 50$ are depicted in Fig. 5. Analysis shows a layered structure with aluminum, cobalt, chromium, iron and nickel constituting the FCC phase and titanium, zirconium, hafnium, niobium and tantalum constituting the BCC phase. The cross-sectional view in the shear direction in Fig. 5a-c and in the radial direction in Fig. 5d-f demonstrates a clear layered structure with the height of layers ranging from the micrometer level to the nanometer level. The thickness of layers is not uniform along the radius due to the fact that the co-deformation of two dissimilar materials, particularly at large shear strain, does not follow an ideal model during HPT processing [43,68]. In fact, with increasing distance from the disc center, the thickness of layers becomes finer, reaching down to about 61 nm, which is much smaller than the one expected from the ideal co-deformation. It should be noted that if the two materials deform ideally, the height of layers after $N$ turns ($h_N$) should be $h_N = h_0 / (1 + 2N) = 7.9$ µm, where $h_0$ is the initial height, which is 800 µm. However, since the layers do not co-deform ideally, the final height of the layers should be much less to keep the rigidity of the material (Fig. 5c), and some vortex-type deformation is also expected (Fig. 5f). It should be emphasized that this deviation from the ideal layer thickness arises from the non-ideal co-deformation involving vortex-type flow and is not related to mechanical alloying. A smaller-than-predicted layer thickness reflects the inadequacy of the ideal geometric co-deformation model for describing the deformation of this material. These results confirm the formation of a nanolamellar structure by HPT processing, while mechanical alloying remains insignificant. Formation of a layered structure by HPT was reported in an earlier study using the Al/Cu system, but mechanical alloying was also achieved due to ultra-fast diffusion [41]. This suggests that the interdiffusion in this system is much slower than in the Al/Cu system. This is not surprising because sluggish diffusion is considered a feature of HEAs [11].

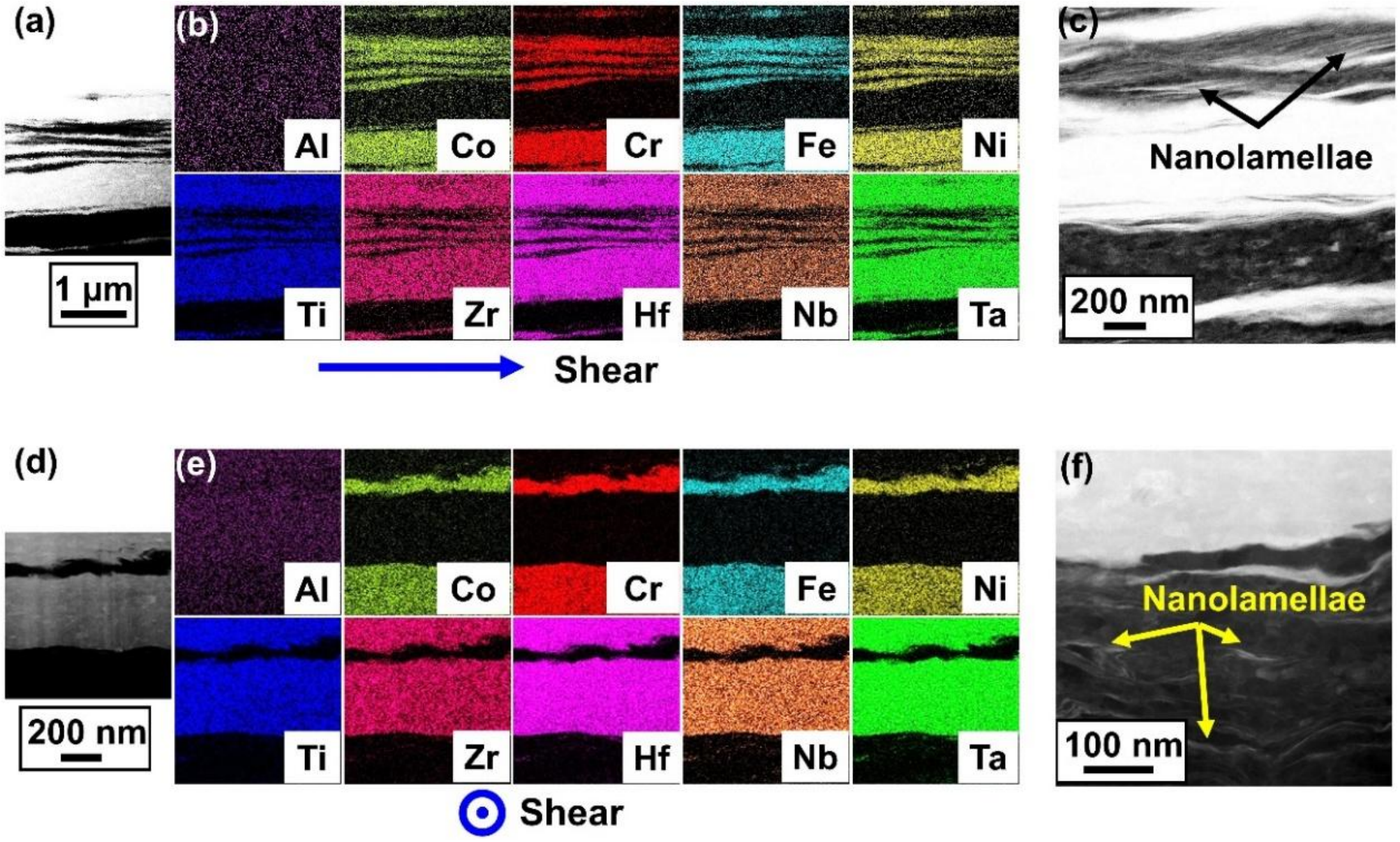

Figure 5. Nanolamellar structure of hybrid high-entropy alloy $Al_{0.1}CoCrFeNi$ + TiZrHfNbTa deformed via high-pressure torsion (HPT). (a, c, d, f) High-angle annular dark-field micrograph and (b, e) elemental maps by energy-dispersive X-ray spectroscopy corresponding to (a) and (d) taken from cross-section in (a-c) shear direction and (d-f) radial direction for specimen deformed with $N = 50$ HPT rotations.

Three-dimensional reconstructed APT elemental mappings of the hybrid material deformed by HPT with $N = 50$ are depicted in Fig. 6a and in a video in the supporting information. The 3D maps show the presence of elements in two phases of $Al_{0.1}CoCrFeNi$ and TiZrHfNbTa, which are elongated in the direction of torsional shear. The main observation in these maps is that the thickness of some layers goes down to nearly 10 nm, confirming the high potential of HPT in fabricating hybrid materials with a nanolamellar structure. A two-dimensional iron compositional map extracted from this volume (Fig. 6b) confirms that some interdiffusion may locally occur where HEA layers are the thinnest. A larger number of HPT turns was avoided in this study, as it was shown earlier that very large strains (i.e. the ultra-SPD process) lead to severe mechanical alloying and formation of new phases rather than the formation of a hybrid material with two distinct phases.

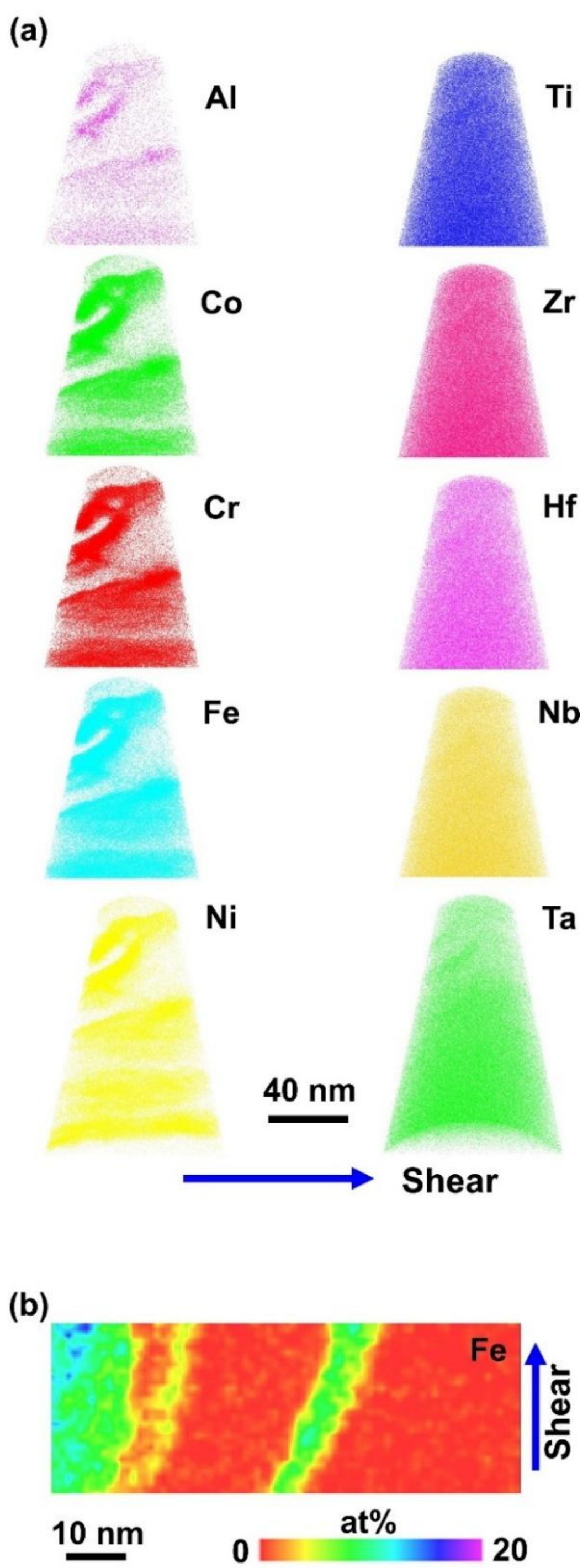

Figure 6. Nanolamellar structure of hybrid high-entropy alloy $Al_{0.1}CoCrFeNi$ + TiZrHfNbTa deformed via high-pressure torsion (HPT). (a) Three-dimensional elemental distribution achieved by atom probe tomography, and (b) two-dimensional compositional map for iron for specimen deformed with $N = 50$ HPT rotations.

Microscopy analysis via TEM is depicted in Fig. 7 for the hybrid material deformed with $N = 50$, where Fig. 7a illustrates a bright-field micrograph, Fig. 7b represents the related SAED pattern, and Fig. 7c presents a dark-field micrograph. The presence of FCC and BCC crystals is verified from the SAED pattern, which is consistent with XRD, EDS and APT analyses. Diffraction rings on SAED patterns indicate the existence of nanosized grains with random misorientations. These nanograins are clearly exhibited in dark-field images. To estimate the level of grain refinement, grain sizes were determined as the average of two orthogonal dimensions of white regions in dark-field images. The grain size distribution (Fig. 7d) shows that the average grain dimension for the mixture of FCC and BCC phases in this hybrid material is 22 nm. This size is similar to that of typical single-phase FCC [69-71] or BCC [19,21] HEAs processed by SPD and significantly smaller than other hybrid alloys [39-42]. It should be noted that this value represents an average grain dimension for both FCC and BCC phases, and it is close to the grain size when individual phases are processed by HPT: 29 nm for $Al_{0.1}$CoCrFeNi [70] and 20 nm for TiZrHfNbTa [21].

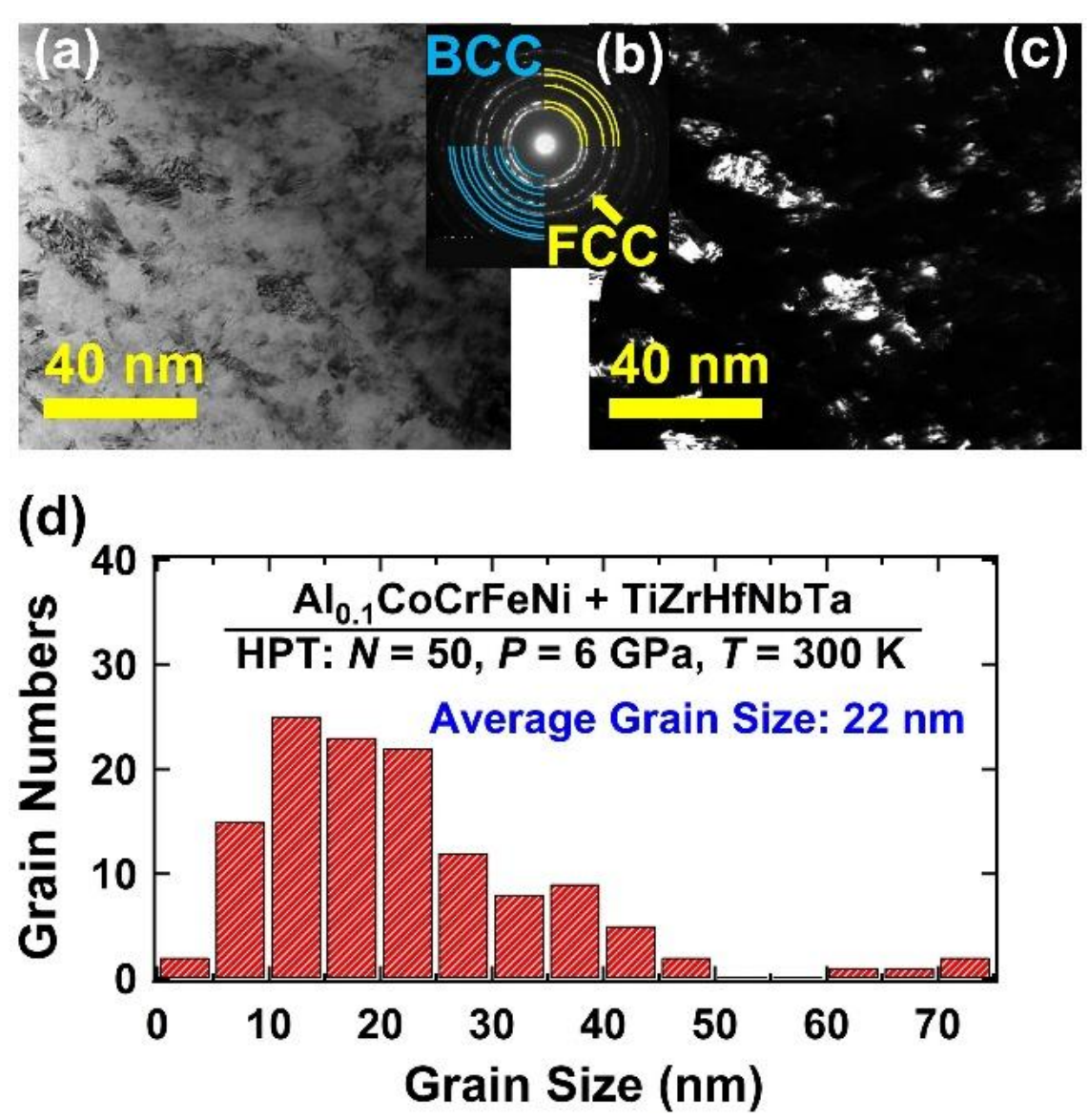

Figure 7. Nanograin formation in hybrid high-entropy alloy $Al_{0.1}$CoCrFeNi + TiZrHfNbTa deformed via high-pressure torsion (HPT). Transmission electron microscopy (a) bright-field micrograph, (b) selected area electron diffraction profile, (c) dark-field image, and (d) histogram of grain size distribution (up disc view) for specimen deformed with $N = 50$ HPT rotations. Dark-field micrograph was obtained using diffraction spots depicted in selected area electron diffraction profile by arrow.

Nanostructural features of the hybrid specimen deformed with $N = 50$ were further analyzed by performing high-resolution TEM. High-resolution TEM images of FCC are depicted in Fig. 8 and 9, and of BCC in Fig. 10. Lattice images of FCC in Fig. 8a show the presence of grain boundaries within a layer of $Al_{0.1}$CoCrFeNi. A closer examination of the interior of FCC grains, as depicted in Fig. 8b, implies that the structure of the FCC phase is distorted by HPT-induced defects. Closer examination of this distorted structure confirms the presence of multiple stacking faults, as shown in Fig. 8c and 8d. Severe strains by HPT should be a major reason for the generation of these stacking faults. Such defects were experimentally detected in some FCC alloys with low stacking fault energies during ball milling [72] and HPT processing [73] and theoretically predicted by molecular dynamics simulations [74]. Analysis of another FCC layer in Fig. 9 confirms the presence of coherent twin boundaries, as shown in Fig. 9a and 9b. Moreover, a low-angle grain boundary formed by dislocation pile-up can be observed in Fig. 9c, and a high-angle grain boundary is depicted in Fig. 9d.

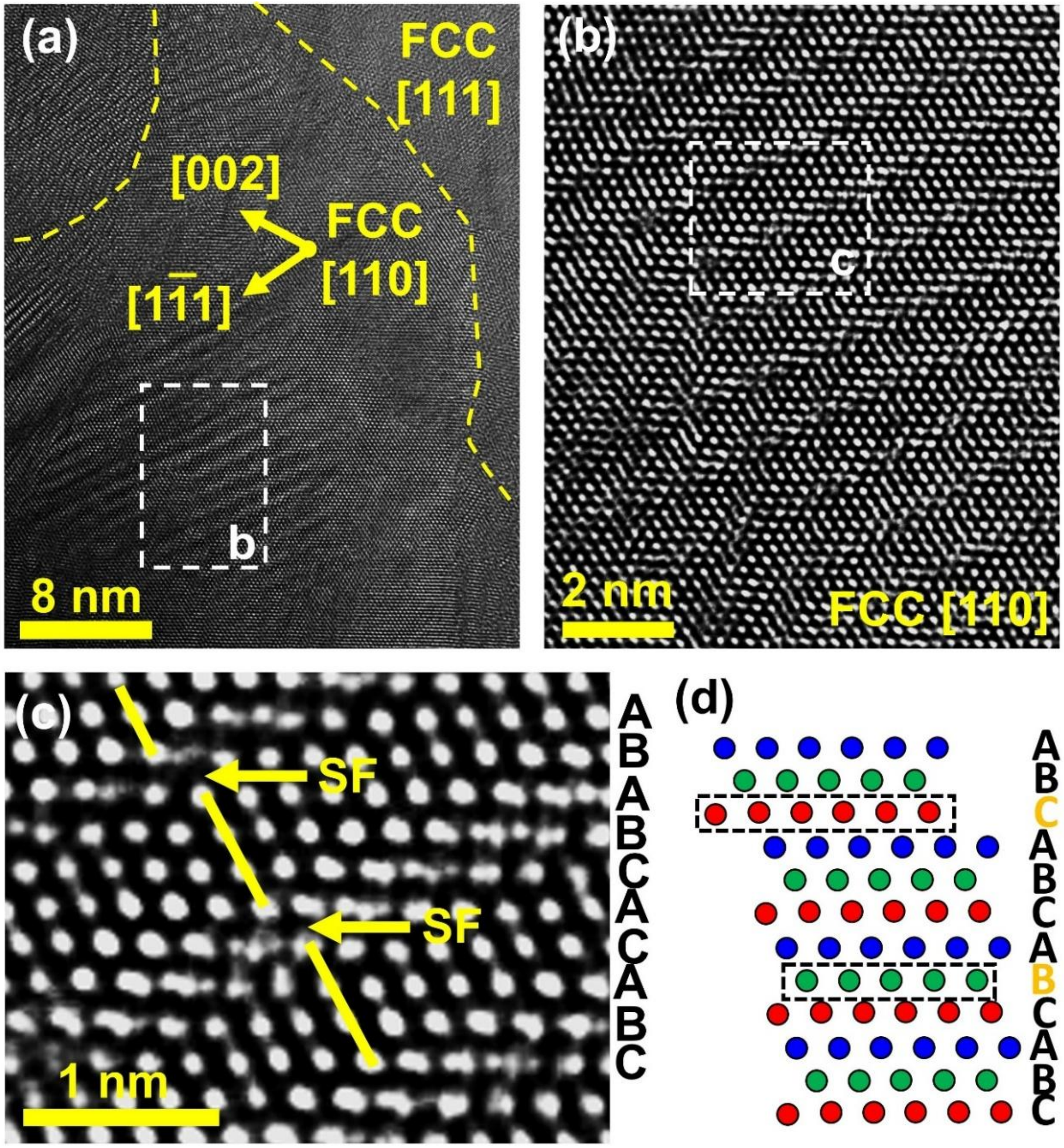

Figure 8. Formation of multiple stacking faults (SF) in FCC structure of hybrid high-entropy alloy $Al_{0.1}$CoCrFeNi + TiZrHfNbTa deformed via high-pressure torsion (HPT). (a) high-resolution transmission electron micrograph, (b, c) lattice images and (d) simulated stacking in FCC for specimen deformed with $N = 50$ HPT rotations. Image (b) is magnification of region indicated in (a) and image (c) is magnification of region depicted in (b).

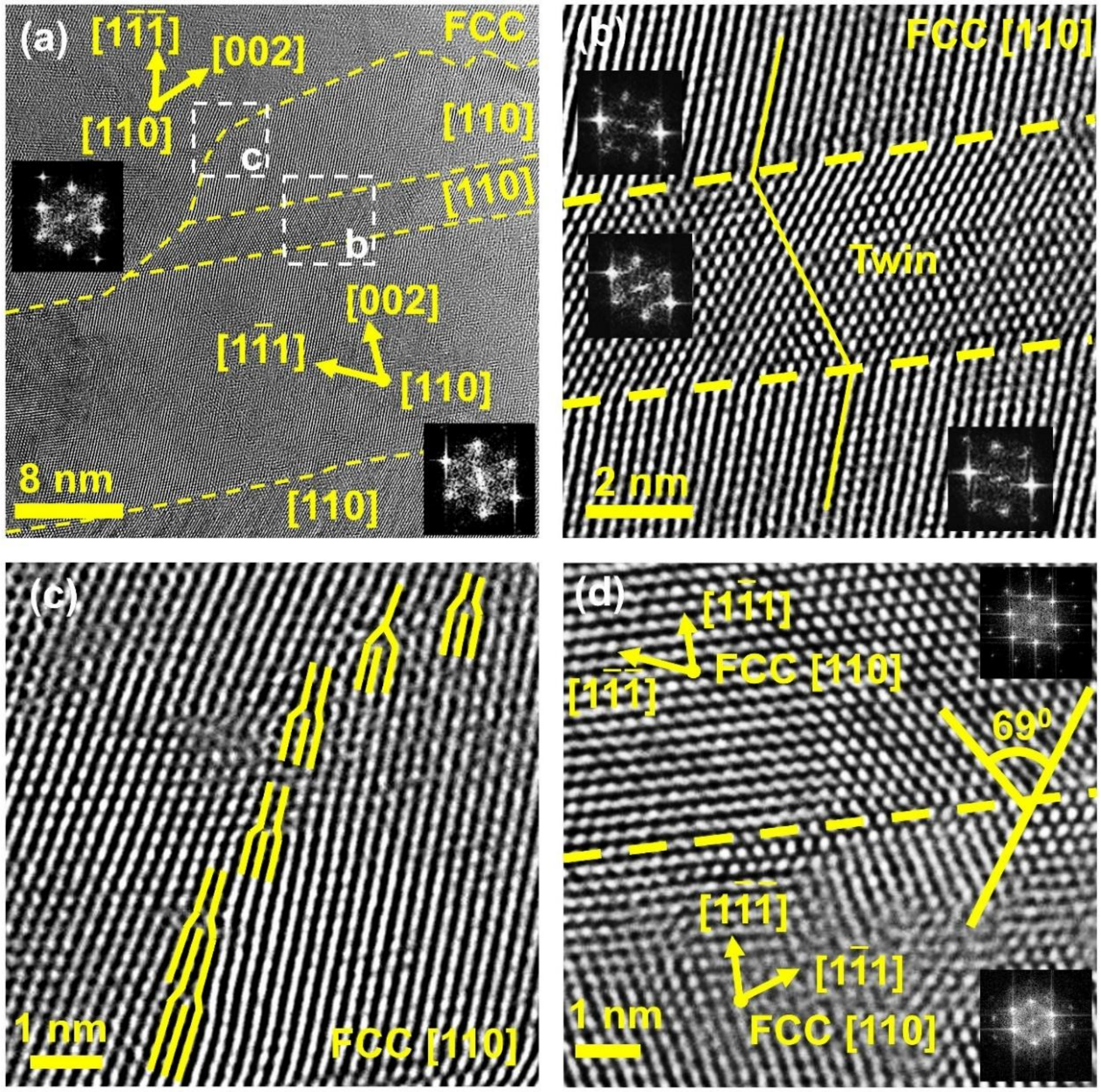

Figure 9. Formation of twin boundaries, dislocation cells and high-angle grain boundaries in FCC structure of hybrid high-entropy alloy $Al_{0.1}CoCrFeNi$ + TiZrHfNbTa deformed with high-pressure torsion (HPT). (a) High-resolution transmission electron micrograph, and lattice images of (b) twins, (c) dislocations and (d) high-angle grain boundary for specimen deformed with $N = 50$ HPT rotations. Images (b) and (c) are magnifications of regions depicted in (a), and insets are fast Fourier transform analysis.

The microstructure of a BCC layer is shown in Fig. 10a, confirming the presence of high-angle grain boundaries between BCC crystals, as depicted in an enhanced magnification in Fig. 10b. Examination of the interior of BCC grains indicates the generation of dislocations, as depicted in Fig. 10c. The stability of dislocations at close distance in this hybrid materials should be due to the impact of multiple solute atoms on distorting the structure and pinning the movement of dislocations. The overall TEM analysis shows that HPT processing not only produces nanolayers of phases, but also generates extreme lattice strain in each layer by the development of stacking faults, grain boundaries, twin boundaries and dislocations. Such defective layered structures are expected to enhance the strength of the material by retaining some ductility [49-51]. The co-presence of two FCC and BCC phases can also contribute to retaining some ductility, as reported in some earlier publications on HEAs [75,76].

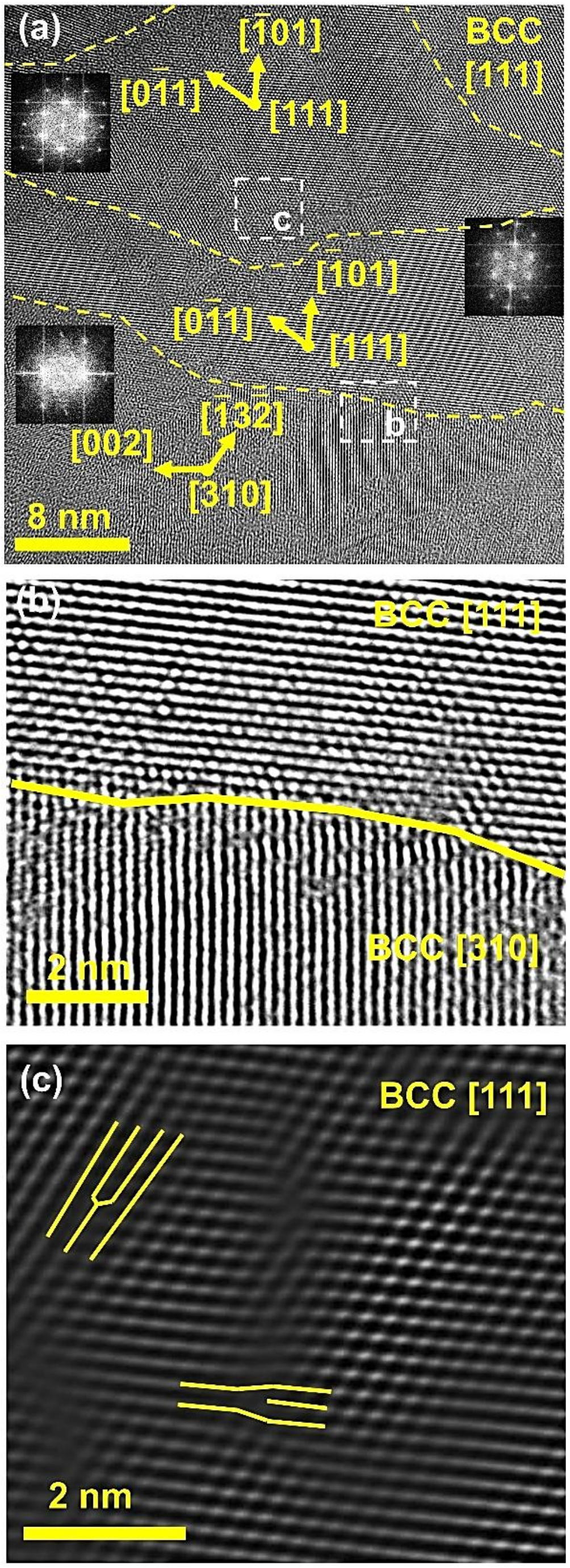

Figure 10. Generation of dislocation and high-angle grain boundaries in BCC structure of hybrid high-entropy alloy $Al_{0.1}CoCrFeNi$ + TiZrHfNbTa deformed via high-pressure torsion (HPT). (a) High-resolution transmission electron micrograph, and lattice images of (b) grain boundary and (c) dislocations for specimen deformed with $N$ = 50 HPT rotations. Images (b) and (c) are magnifications of regions indicated in (a), insets are fast Fourier transform (FFT) analysis, and image (c) was reconstructed by inverse FFT.

Microhardness of the specimens deformed with $N = 1$, 10 and 50 is depicted in Fig. 11a versus the radial locations from the specimen center. An increase in microhardness is observed with the HPT turns and the radial distance. The highest microhardness value of 740 Hv is achieved after $N = 50$ turns at a radial distance of 4 mm. The hardness in this hybrid material is higher than the constituent HEA $Al_{0.1}CoCrFeNi$ after HPT processing [70,71], TiZrHfNbTa after HPT processing [21,63] and other HPT-processed composites [39-42]. The reason for a higher hardness in this investigation can be ascribed to the combined effect of lattice defects [23-25] and nanolamellar formation [77,78]. Variation of hardness in Fig. 11a is replotted versus the shear strain in Fig. 11b, where shear strain was estimated through $\gamma = 2\pi rN/hr$, in which $r$, $N$ and $h$ are radial distance, HPT turns and specimen thickness, respectively [26,27]. A strain hardening effect is observed with increasing shear strain; however, the hardening rate diminishes at higher shear strain. The tendency of hardness and microstructure to saturate at the steady states is caused by the activation of dynamic recovery [79,80], recrystallization [81,82] and grain-boundary migration [81,82]. For further details about the mechanism of hardness and microstructure evolution with strain and the occurrence of saturation, the reader is referred to a recent review paper [83].

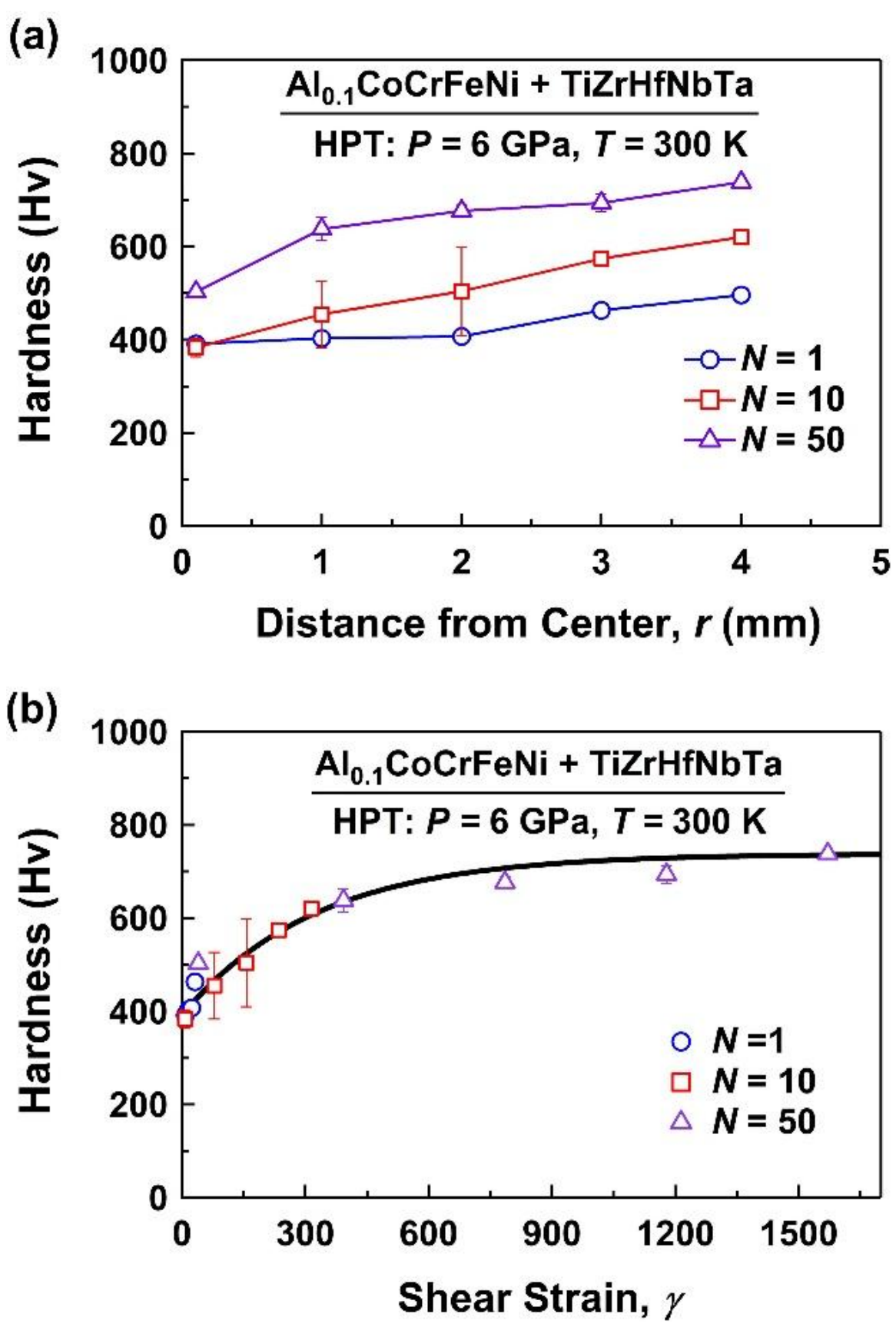

Figure 11. High microhardness of hybrid high-entropy alloy $Al_{0.1}CoCrFeNi$ + TiZrHfNbTa deformed via high-pressure torsion (HPT). Variations of microhardness as a function of (a) radial distance on disc and (b) shear strain for specimens deformed with $N = 1$, 10 and 50 HPT rotations.

Nominal stress-strain curve achieved via micro-tensile testing of hybrid material deformed with $N = 50$ is depicted in Fig. 12a. The material exhibits an ultimate tensile strength of 2.4 GPa.

Despite such a high strength, the material exhibits some plasticity before failure. Quantitatively, the hybrid material deformed with $N = 50$ shows a uniform plastic strain of about 2% up to the ultimate tensile strength and a total elongation of about 4% at fracture. Fig. 12b-e illustrates SEM micrographs of the sample after the fracture. The overall views in Fig. 12b and 12c show that the surface of the samples remains smooth due to low plastic deformation, in agreement with the stress-strain curve. However, a closer inspection indicates the formation of some shear lines on the surface, as shown in Fig. 12d, and some dimples, as depicted in Fig. 12e. These observations imply that the material can still accommodate some plasticity despite its ultrahigh strength. It should be noted that the as-processed structure already contains a high density of defects introduced by HPT, including dislocations, stacking faults, and twin boundaries, which are themselves the primary carriers of plasticity in this material. The evidence for plastic deformation is therefore based on the fractographic observations after tensile and bending testing. Bending stress versus the bending displacement is shown in Fig. 13a. The material shows a maximum bending strength of 4.0 GPa with ductile behavior. SEM micrographs of deformation morphology after the bending test, as shown in Fig. 13b and 13c confirm a large plastic deformation of the sample. To examine the reliability of the micromechanical data, another bending test was conducted on a sample with $10\times10\times40$ $\mu m^3$ dimensions, which resulted in a bending strength of 4.5 GPa, confirming reasonable reproducibility of the tests. It should be noted that although Euler-Bernoulli beam theory assumes linear elasticity, the measured ratio of bending strength to tensile strength (4.0 GPa / 2.4 GPa = 1.67) is consistent with the theoretical shape factor for a beam in pure bending, which ranges from 1.5 for a rectangular cross-section to 1.7 for a circular cross-section [84]. This ratio agreement, which was observed in earlier works of the current authors on HPT-processed titanium, hafnium and a Nb-Ti alloy. suggests that the calculated bending strength is physically valid. These ultrahigh strength and reasonable plasticity originate from the combination of nanolayered structure [85] with a large portion of grain boundaries [86], twin boundaries [5], dislocations [36] and stacking faults [87].

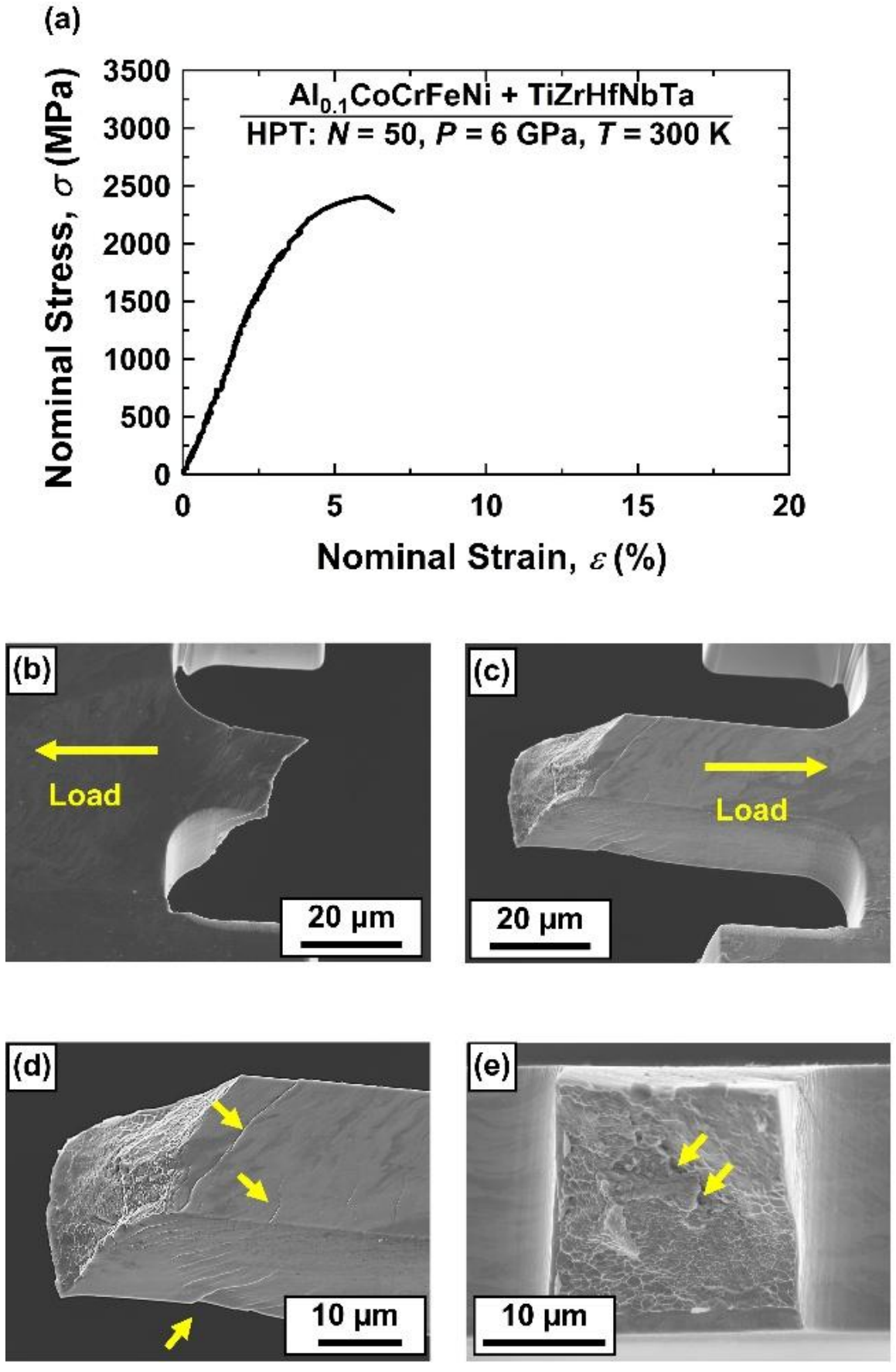

Figure 12. High tensile strength with reasonable plasticity for hybrid high-entropy alloy $Al_{0.1}CoCrFeNi$ + TiZrHfNbTa deformed via high-pressure torsion (HPT). (a) Micro-tensile testing stress-strain curve, (b-e) scanning electron micrographs of tensile samples after fracture of specimens deformed with $N$ = 50 HPT rotations. Image (d) is magnification of (c) with shear features indicated by arrows, and image (e) is fracture surface with dimples indicated by arrows.

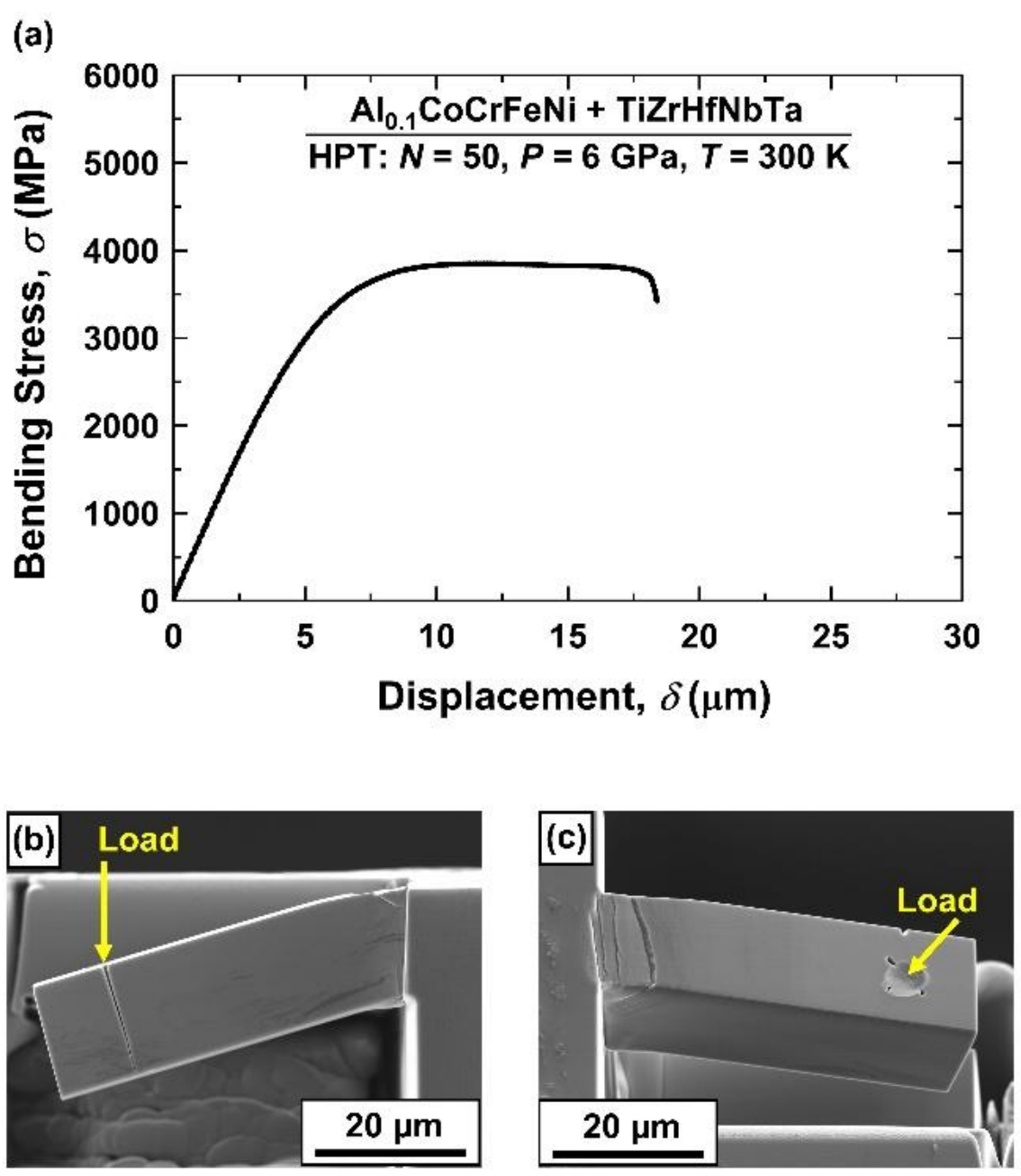

Figure 13. High bending strength with reasonable ductility for hybrid high-entropy alloy $Al_{0.1}$CoCrFeNi + TiZrHfNbTa deformed via high-pressure torsion (HPT). (a) Bending stress versus bending displacement, (b) side view and (c) top view of micro-bending specimen taken by scanning electron microscopy after bending test for specimen deformed with $N = 50$ HPT rotations.

**Discussion**

This study introduces a nanolamellar hybrid HEA material by processing via HPT. After processing by 50 turns of HPT, the dual-phase material with a nanolamellar structure demonstrates a hardness of 740 Hv, an ultimate tensile strength of 2.4 GPa and a maximum bending strength of 4.0 GPa with some ductility/plasticity. These remarkable properties demonstrate the ability of the hybrid nanolamellar high-entropy materials to beat the inverse strength and plasticity relation. Three questions in this research need to be further discussed: (i) the formation mechanism of defect-rich layered nanostructure, (ii) parameters responsible for high strength and reasonable ductility, and (iii) how the mechanical properties of hybrid HEA $Al_{0.1}$CoCrFeNi + TiZrHfNbTa stand relative to previously reported hybrid or HEA materials.

The formation of layered microstructure and nanograins is due to the effect of SPD on the alloy through HPT [26]. HPT subjects the alloy to extreme pressures of several gigapascals with simultaneous torsional straining, thereby causing not only an extreme grain refinement [27] but also forming a layered structure by co-deformation of two dissimilar materials [39-42]. During the co-deformation process, initially, the equiaxed grains get aligned in the shear direction, as found in this study by SEM analysis. With increasing shear strains, the elongated phases become thinner down to the nanometer level by shearing, giving rise locally to some intermixing [39-42], as observed by STEM and APT in this study. This intermixing with an alternate nanolamellar structure may reduce dynamic recovery processes [79,80] because dislocations (stacking faults and twins in FCC [73]) are necessary to accommodate distortions resulting from co-deformation. Such a structure also hinders dynamic recrystallization [81,82] and grain boundary migration [81,82] because interphase boundaries exhibit low mobility. As a result, not only a nanolamellar structure is formed, but also large fractions of grain boundaries, twin boundaries, dislocations, and stacking faults are generated within each layer. The thickness of these nanolamellar structures and grain size could be further reduced by increasing the shear deformation by HPT, but this was avoided in this study due to the risk of mechanical alloying by ultra-SPD [43]. It should be noted that although XRD and SEM-EDS suggest that overall atomic-scale mechanical alloying between the two phases is negligible, some localized intermixing was detected by APT, particularly close to interphase boundaries.

Regarding the combination of high strength and reasonable ductility/plasticity achieved in the nanolamellar HEA structure, it obviously results from the combination of several mechanisms. The high strength in SPD-deformed specimens mainly originates from the presence of nanograins [26,27], as the grain boundaries restrict the movement of dislocations via the Hall-Petch mechanism [1,3]. Furthermore, the presence of twin boundaries [5], stacking faults [87] and dislocations [80] can influence the overall strength of the specimen, although it is hard to quantify the influence of each kind of defect on the overall strength because of a lack of analytical information for HEAs. However, it is possible to show that grain boundaries and defects are not sufficient to achieve such a high hardness and strength. This hybrid material with equal volume fractions of BCC and FCC phases and a 22 nm grain size exhibits a microhardness of 740 Hv, against 520 Hv for the HPT-processed $Al_{0.1}CoCrFeNi$ HEA (average grain size of 29 nm) [70] and 500 Hv for the HPT-processed TiZrHfNbTa HEA (average grain size of 20 nm) [21]. It should be noted that FCC and BCC phases in Refs. [21,70] were processed using the same machine, under similar HPT conditions and examined by the same microhardness testing as in this study, except that the number of turns was $N = 10$ (grain size reached a steady state after $N = 2$). If it is reasonably assumed that the mean size of grains and, accordingly, defect density in this hybrid compound is close to the HPT-processed HEAs (quantification of defects in these strained nanostructured materials is challenging), the rule of mixtures can be reasonably employed to estimate the microhardness of the hybrid materials. The rule of mixtures leads to a value of 510 Hv, which is 70% of the experimental value for the hybrid material (740 Hv). Thus, about 30% hardening beyond the Hall-Petch and defect-induced hardening originates from the particular nanolamellar structure of this material.

A full quantitative separation of the interphase, grain-boundary and defect contributions would require comprehensive studies on quantification of major strengthening mechanisms in single-phase HEAs with and without mechanical processing, followed by modification of such quantifications to dual-phase HEA hybrid materials and finally by extending them to nanolamellar hybrid HEAs. Therefore, the value of 30% should be regarded as an estimation rather than a precise, directly measured quantity, given the assumptions inherent in applying the rule of mixtures to strained nanostructured materials. The presence of a layered interface structure in composites impedes the movement of dislocations through the interphase/interface boundary hardening mechanism [75,76]. Such interface-controlled hardening, which depends on interface coherency, interface structure, and the mechanical properties and slip systems of the two phases, does not directly appear in the Hall-Petch relationship. Similar interface-controlled hardening, which can be due to various factors such as internal stress and localized dislocation accumulation due to the difference in the deformation of the two phases, has been reported in composites [88,89] and heterostructured materials [90,91], and it is particularly significant in nanolamellar structures [92]. The local atomic intermixing near interphase boundaries (Fig. 6b) can potentially affect the interphase hardening phenomenon, although clarification of the significance of this effect remains a subject for future interface-specific mechanical testing or atomistic simulation. It should be noted that texture evolution during HPT deformation can also influence strengthening behavior. Although this issue was not evaluated deeply in this study (XRD does not suggest preferential texture in Fig. 2b), texture is not usually significant when metals with cubic phases like BCC or FCC are processed by HPT [93].

The presence of some ductility despite the ultrahigh strength of this hybrid material can be primarily explained by the heterostructured nature of the material. Although in heterostructured materials, the co-presence of soft and hard phases is needed [27], the co-deformation of two different phases can also generate positive stress on one phase and a negative stress on the other phase because two phases exhibit different flow stress. Thus, the phases have a large level of internal strain, particularly at the nanoscale, which allows them to accommodate some dislocations to overcome such stresses, similar to pearlitic steels [75,76]. Twin boundaries, observed in the FCC phase in this study, were also suggested to accommodate dislocations [5,6], a fact that was also reported in HEAs [94,95]. The presence of stacking faults in FCC can be another reason for retaining some plasticity, as it was shown earlier that structures with dense stacking faults tend to accommodate dislocations to release the stress in hybrid materials [87].

To address the third question, a direct comparison with reported data in the literature is shown in Table 2 [21,44-48,85,96-113] and Fig. 14. Hybrid HEA $Al_{0.1}CoCrFeNi$ + TiZrHfNbTa processed by $N = 50$ turns of HPT achieves the highest ultimate tensile strength among all HEAs and metallic composites, reaching 2.4 GPa, although its elongation failure is not as high as other materials due to the trade-off relationship between strength and ductility. Moreover, the hybrid material also exceeds the tensile strength of both of its individually HPT-processed constituent alloys. These results establish nanolamellar hybrid HEAs as potential candidates for applications when a high strength is desirable. Moreover, this study re-confirms the potential of HPT in developing advanced nanocomposites, a topic that has been extended even to ceramic composites in recent years [114,115].

Table 2. Ultimate tensile strength and elongation at fracture of hybrid high-entropy alloy $Al_{0.1}CoCrFeNi$ + TiZrHfNbTa deformed via $N = 50$ high-pressure torsion (HPT) turns compared with previously reported hybrid and high-entropy alloys processed by HPT, accumulative-roll bonding (ARB), equal-channel angular pressing (ECAP), cold rolling (CR), laser powder bed fusion (L-PBF), laser-based directed energy deposition (L-DED), and directional solidification (DS).

| **Materials** | **Processing** | **Ultimate Tensile Strength (GPa)** | **Elongation at Fracture (%)** | **Reference** |
|---|---|---|---|---|
| $Al_{0.1}CoCrFeNi$+TiZrHfNbTa | HPT | 2.40 | 4 | This study |
| TiZrHfNbTa | HPT | 2.13 | 13 | [21] |
| Cu+Ni+Al | ARB | 0.26 | 1 | [44] |
| Al+Cu+Sn+Ni | ARB | 0.14 | 7 | [45] |
| Al 1060+Al 7N01 | ARB | 0.26 | 7 | [46] |
| Al+Cu+Zn+Ni | ARB | 0.31 | 12 | [47] |
| Cu/CuAg+NbMoTaW | ARB | 0.56 | 15 | [48] |
| $Al_{19}Co_{20}Fe_{20}Ni_{41}$ (Eutectic) | L-PBF | 1.63 | 22 | [7] |
| $Al_{0.1}CoCrFeNi$ | Twisting | 1.06 | 22 | [96] |
| TiZrHfNbTa | HPT | 1.64 | 13 | [97] |
| $AlCoCrFeNi_{2.1}$ (Eutectic) | L-PBF | 1.38 | 19 | [98] |
| $AlCoCrFeNi_{2.1}$ (Eutectic) | CR | 1.56 | 13 | [99] |
| $AlCoCrFeNi_{2.1}$ (Eutectic) | L-PBF | 1.43 | 20 | [100] |
| $Al_{19.4}Fe_{18.5}Co_{18.5}Ni_{43.6}$ (Eutectic) | L-DED | 1.41 | 23 | [101] |
| $AlCoCrFeNi_{2.1}$ (Eutectic) | L-DED | 1.25 | 23 | [102] |
| $Ni_{43.9}Co_{17}Fe_{12}Cr_{9}Al_{18}B_{0.1}$ (Eutectic) | Rolling | 1.3 | 24 | [103] |
| $Al_{1.19}CoFeNi_{2.86}$ (Eutectic) | DS | 1.01 | 10 | [104] |
| CoCrFeMnNi | ARB | 1.29 | 22 | [105] |
| FeCoCrNiMn | L-PBF | 0.91 | 23 | [106] |
| $Fe_{39}Ni_{37}Cr_{8}Si_{8}A_{18}$ | CR | 1.82 | 15 | [107] |
| $Al_{1.25}CoCrFeNi_{2.8}Mo_{0.2}$ (Eutectic) | DS | 1.17 | 19 | [108] |
| CoCrFeNi+$Al_{0.3}CoCrFeNi$ | Hot Pressing + CR | 1.32 | 26 | [109] |
| FeCoNiMnV | Homogenization | 0.47 | 34 | [110] |
| FeCoNiMnV | CR | 0.72 | 15 | [110] |
| $Al_{0.25}FeCoNiMnV$ | Homogenization | 0.59 | 37 | [110] |
| $Al_{0.25}FeCoNiMnV$ | CR | 0.96 | 18 | [110] |
| $Ti_{0.25}FeCoNiMnV$ | Homogenization | 0.75 | 36 | [110] |
| $Ti_{0.25}FeCoNiMnV$ | CR | 1.12 | 26 | [110] |
| CoCrFeMnNi+304L Steel | CR | 1.34 | 9 | [111] |
| CoCrFeMnNi+304L Steel | CR + Annealing (898 K) | 0.73 | 27 | [111] |
| CoCrFeMnNi+304L Steel | CR + Annealing (1048 K) | 0.44 | 45 | [111] |
| CoCrNi | ECAP | 1.50 | 4 | [112] |
| $Al_{0.1}CoCrFeNi$ | ECAP | 0.74 | 16 | [113] |
| $Al_{0.1}CoCrFeNi$ | Casting | 0.50 | 54 | [113] |
| $Al_{0.1}CoCrFeNi$ | ECAP + Annealing (773 K) | 0.68 | 18 | [113] |
| $Al_{0.1}CoCrFeNi$ | ECAP + Annealing (973 K) | 0.60 | 25 | [113] |
| $Al_{0.1}CoCrFeNi$ | ECAP + Annealing (1173 K) | 0.59 | 42 | [113] |

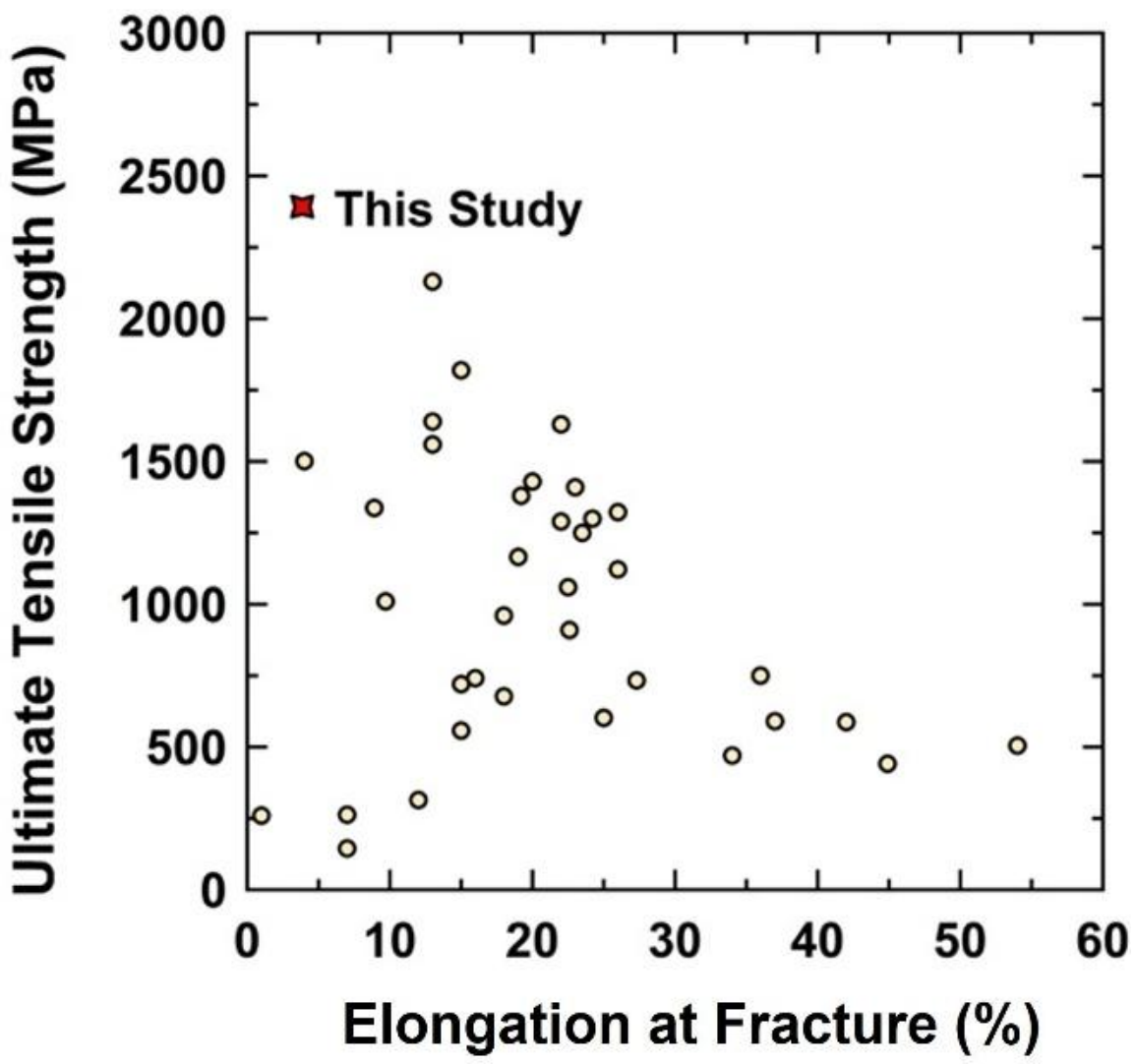

Figure 14. Ultimate tensile strength versus elongation at fracture for hybrid high-entropy alloy $Al_{0.1}CoCrFeNi$ + TiZrHfNbTa deformed via $N$ = 50 high-pressure torsion (HPT) turns compared with reported data in the literature, which is given in Table 2.

## Conclusions

A hybrid $Al_{0.1}CoCrFeNi$ + TiZrHfNbTa compound is designed by deformation via high-pressure torsion. The hybrid material has a dual-phase nanolamellar structure consisting of face-centered cubic and body-centered cubic phases. Within each phase layer, large fractions of defects like grain boundaries, dislocations, twin boundaries and stacking faults are present. The compound exhibits remarkable properties, demonstrating a hardness of 740 Hv, a tensile strength of 2.4 GPa and a bending strength of 4.0 GPa with some ductility under both tensile and bending loads. This study introduces hybrid dual-phase high-entropy alloys with a nanolamellar configuration as new candidates for ultrahigh-strength applications.

## Data availability

Data will be made available on request.

## CRediT Authorship Contribution Statement

All authors: Conceptualization, Methodology, Investigation, Validation, Writing – review & editing.

## Declaration of Competing Interest

The authors declare no competing financial interests or personal relations that could affect the work presented in the current manuscript.

## Acknowledgment

The author S.D. is thankful to the MEXT, Japan, for a scholarship. This study was supported in part by the Light Metals Educational Foundation of Japan, in part by the Japan Science and Technology Agency (ASPIRE JPMJAP2332), and in part by the CNRS Federation IRMA - FR 3095 and the Region Normandy. The authors acknowledge Diamond Light Source for access to the I12-JEEP beamline for synchrotron diffraction under project number MG38691.

## Supplementary material

Supplementary material of this article contains a video of the nanostructure of the material constructed by APT.